\documentclass[a4paper,12pt]{article}

\usepackage{graphicx}
\usepackage[total={6in,9in},top=1in,left=1.25in]{geometry}

\usepackage[
  all=tight,  
  mathspacing=normal, mathdisplays=normal, 
  leading=normal, 
  margins=normal  
  ]{savetrees}

\newif\ifdraft

\usepackage{amsmath,amssymb}

\usepackage{shadethm}
\AtBeginDocument{%
\definecolor{shadethmcolor}{gray}{1}
\shadeboxrule0pt
}

\newshadetheorem{thm}{\bfseries\sffamily Theorem}[section]
\newshadetheorem{prp}[thm]{\bfseries\sffamily Proposition}
\newshadetheorem{lem}[thm]{\bfseries\sffamily Lemma}
\newshadetheorem{cor}[thm]{\bfseries\sffamily Corollary}
\newshadetheorem{clm}[thm]{\bfseries\sffamily Claim}
\newshadetheorem{dfn}[thm]{\bfseries\sffamily Definition}
\newshadetheorem{ex}[thm]{\bfseries\sffamily Example}
\newshadetheorem{asm}[thm]{\bfseries\sffamily Assumption}
\newshadetheorem{obs}[thm]{\bfseries\sffamily Observation}

\newenvironment{prf}[1][Proof]{\begin{trivlist}
\item[\hskip \labelsep {\bfseries\sffamily #1.}]}{\end{trivlist}}
\newcommand\qed{\ensuremath\square}

\makeatletter
\usepackage[\ifdraft draft,\fi
  colorlinks=true,
  breaklinks
]{hyperref}
\hypersetup{allcolors=[rgb]{.0,.0,.67}}
\renewcommand\eqref[1]{\hyperref[#1]{\text{(\ref*{#1})}}}
\newcommand\secref[1]{\hyperref[#1]{Section~\ref*{#1}}}
\newcommand\figref[1]{\hyperref[#1]{Figure~\ref*{#1}}}
\newcommand\tabref[1]{\hyperref[#1]{Table~\ref*{#1}}}
\newcommand\exref[1]{\hyperref[#1]{Example~\ref*{#1}}}
\newcommand\dfnref[1]{\hyperref[#1]{Definition~\ref*{#1}}}
\newcommand\chref[1]{\hyperref[#1]{Chapter~\ref*{#1}}}
\newcommand\asmref[1]{\hyperref[#1]{Assumption~\ref*{#1}}}
\newcommand\lemref[1]{\hyperref[#1]{Lemma~\ref*{#1}}}
\newcommand\thmref[1]{\hyperref[#1]{Theorem~\ref*{#1}}}
\newcommand\corref[1]{\hyperref[#1]{Corollary~\ref*{#1}}}
\newcommand\prpref[1]{\hyperref[#1]{Proposition~\ref*{#1}}}
\newcommand\algref[1]{\hyperref[#1]{Algorithm~\ref*{#1}}}
\newcommand\nref[2][]{\hyperref[#2]{\text{#1\ref*{#2}}}}
\makeatother

\usepackage[ruled,linesnumbered]{algorithm2e}
\SetCommentSty{textit}
\SetDataSty{texttt}
\SetFuncSty{texttt}
\SetProcFnt{\small}
\SetProcNameFnt{\small}
\SetProcNameSty{texttt}
\SetProcArgFnt{\small}
\SetAlFnt{\footnotesize}
\SetAlCapFnt{\bfseries\sffamily}
\newcommand\hhrule{\hrule height .75pt}

\newcommand\block[1]{\par\noindent\makebox{\textit{#1}}}

\newsavebox\tmpbox
\newcounter{enum}
\newcommand\labelenum{\upshape\theenum}
\newlength\enumlmargin
\newenvironment{enum}{%
  \begin{list}{\labelenum}{%
    \setcounter{enum}1%
    \sbox\tmpbox{\labelenum}%
    \setlength\labelwidth{\wd\tmpbox}%
    \setlength\leftmargin\labelwidth%
    \addtolength\leftmargin\labelsep%
    \addtolength\leftmargin\enumlmargin%
    \setlength\topsep{0pt}%
    \setlength\partopsep{0pt}%
    \setlength\itemsep{0pt}%
    \setlength\parsep{0pt}%
    \usecounter{enum}%
  }}
  {\end{list}}

\newcounter{enumcasesi}
\newenvironment{enumcases}{%
  \setcounter{enumcasesi}0%
  \setlength\itemsep{1pt plus 2pt minus 1pt}%
  \setlength\topsep\itemsep%
  \setlength\partopsep{0pt}%
  \begin{trivlist}%
  \let\olditem\item%
  \renewcommand\item[1]{\olditem[\refstepcounter{enumcasesi}\hskip\labelsep\itshape Case \theenumcasesi: ##1]}}
  {\end{trivlist}}
  
\newcommand\st{s.\,t.}
\newcommand\ie{i.\,e.}
\newcommand\eg{e.\,g.}

\newcommand\lhs{LHS}
\newcommand\rhs{RHS}
\newcommand\wrt{w.\,r.\,t.}

\newcommand\C{\mathcal C}

\renewcommand\P{\mathcal P} 

\renewcommand\S{\mathcal S} 
\newcommand\T{\mathcal T}

\newcommand\RR{\mathbb R}
\newcommand\X{\mathcal X}
\newcommand\eps{\varepsilon}

\newcommand\writealg[1]{\textsf{\smaller\upshape#1}}
\newcommand\newalg[2]{\expandafter\newcommand\csname#1\endcsname{\writealg{#2}}}

\allowdisplaybreaks

\def\(#1\){\begin{align*}#1\end{align*}}
\newcommand\eqlabel[1]{\stepcounter{equation}\tag{\theequation}\label{#1}}

\makeatletter
\def\@mathclap#1#2{\hbox to 0pt{\hss{$\mathsurround=0pt#1{#2}$}\hss}}
\def\mathclap{\mathpalette\@mathclap}
\makeatother

\mathchardef\ordinarycolon\mathcode`\:
\mathcode`\:=\string"8000
\begingroup \catcode`\:=\active
  \gdef:{\mathrel{\mathop\ordinarycolon}}
\endgroup

\renewcommand\leq\leqslant
\renewcommand\geq\geqslant

\makeatletter
\def\Hv@scale{.95}  
\makeatother

\makeatletter

\def\capstyle{\sffamily\bfseries} 
\long\def\@makecaption#1#2{%
  \vskip\abovecaptionskip
  \sbox\@tempboxa{\small{\capstyle#1:} #2}%
  \ifdim \wd\@tempboxa >.96\hsize
    \centering\parbox{.96\hsize}{\small{\capstyle#1:} #2}
  \else
    \global \@minipagefalse
    \hb@xt@\hsize{\hfil\box\@tempboxa\hfil}%
  \fi
  \vskip\belowcaptionskip}

\def\chapstyle{\rmfamily\scshape} 
\def\parsecstyle{\sffamily\bfseries}  

\def\@seccntformat#1{\csname the#1\endcsname\enskip}

\def\section{\@startsection{section}{1}{\z@}%
  {-3.25ex \@plus -1ex \@minus -.2ex}%
  {1.3ex \@plus.2ex}%
  {\raggedright\large\parsecstyle}}
\def\subsection{\@startsection{subsection}{2}{\z@}%
  {-3ex\@plus -1ex \@minus -.2ex}%
  {.8ex \@plus .2ex}%
  {\raggedright\large\parsecstyle}}
\let\subsubsection\undefined
\def\paragraph{\@startsection{paragraph}{4}{\z@}%
  {\parskip}%
  {-1em}%
  {\raggedright\normalsize\parsecstyle}}

\def\section{\@startsection{section}{1}{\z@}%
  {-1.5ex \@plus -.5ex \@minus -.2ex}%
  {.5ex \@plus.2ex}%
  {\raggedright\large\parsecstyle}}
\def\subsection{\@startsection{subsection}{2}{\z@}%
  {-1.25ex\@plus -.5ex \@minus -.2ex}%
  {.5ex \@plus .2ex}%
  {\raggedright\large\parsecstyle}}
\let\subsubsection\undefined
\def\paragraph{\@startsection{paragraph}{4}{\z@}%
  {\z@}%
  {-.5em}%
  {\raggedright\normalsize\parsecstyle}}
  
\def\@makechapterhead#1{%
  {\vspace*{40\p@}
  \parskip1.5ex
  \parindent\z@
  \raggedright
  \ifnum \c@secnumdepth >\m@ne%
    \if@mainmatter%
      {\chapstyle\large\@chapapp~\thechapter}\par\nobreak
    \fi%
  \fi%
  \interlinepenalty\@M%
  {\parsecstyle\Huge #1}\par\nobreak
  \vskip 30\p@
  }}
\def\@makeschapterhead#1{%
  {\vspace*{40\p@}
  \parskip1.5ex
  \parindent\z@
  \raggedright
  \interlinepenalty\@M%
  {\parsecstyle\Huge #1}\par\nobreak
  \vskip 30\p@
  }}

\def\ps@thesis{%
  \let\@oddfoot\@empty\let\@evenfoot\@empty%
  \def\@evenhead{\vbox to 12pt{\raggedright\thepage\hspace{.5em}\leftmark\hfil\vfil\hhrule}}%
  \def\@oddhead{\vbox to 12pt{\hfil\raggedleft\rightmark\hspace{.5em}\thepage\vfil\hhrule}}%
  \let\@mkboth\markboth
\def\chaptermark##1{%
  \markboth {
	\ifnum \c@secnumdepth >\m@ne
	  \if@mainmatter
		{\chapstyle\@chapapp\space\thechapter.\space}
	  \fi
	\fi
	\sffamily##1}{}}%
\def\sectionmark##1{%
  \markright {%
	{\parsecstyle\mdseries
	\ifnum \c@secnumdepth >\z@
	  {\thesection.\space}%
	\fi
	##1}}}
}

\makeatother

\newif\iffullpaper
\fullpapertrue 

\AtBeginDocument{%
\abovedisplayskip 7pt plus3pt minus 2pt
\belowdisplayskip 7pt plus3pt minus 2pt
\abovedisplayshortskip 0pt plus 3pt
\belowdisplayshortskip 0pt plus 3pt
}

\renewcommand\T{\mathcal T} 

\newalg{update}{Update}
\newalg{pws}{PWS}
\newalg{ps}{PS}
\newalg{mdl}{MDL}

\title{\Large\parsecstyle Generalized Probability Smoothing}
\author{%
{\small Christopher Mattern}\\
{\small DeepMind}\\
{\small London, United Kingdom}\\
{\small \texttt{cmattern@google.com}}
}
\date{}
\begin{document}

\thispagestyle{empty}


\thispagestyle{empty} 

{%
\linespread{.8}
\maketitle
}

\vbox{%
\centering%
\parbox{.8\linewidth}{%
\small%
\paragraph{\small Abstract.}
In this work we consider a generalized version of Probability Smoothing, the core elementary model for sequential prediction in the state of the art PAQ family of data compression algorithms.
Our main contribution is a code length analysis that considers the redundancy of Probability Smoothing with respect to a Piecewise Stationary Source.
The analysis holds for a finite alphabet and expresses redundancy in terms of the total variation in probability mass of the stationary distributions of a Piecewise Stationary Source.
By choosing parameters appropriately Probability Smoothing has redundancy $O(S\cdot\sqrt{T\log T})$ for sequences of length $T$ with respect to a Piecewise Stationary Source with $S$ segments.
}}
\vspace{10pt}


\section{Introduction}

\paragraph{Background.}
Online probability estimation is a central building block of every statistical data compression algorithm \cite{hbdc}, \eg\ Prediction by Partial Matching, Context Tree Weighting and PAQ (pronounced ``pack'').
Statistical data compression involves two phases, modeling and coding, and these phases apply to an input sequence letter by letter.
A statistical \emph{model} predicts a distribution $p$ on the possible outcomes of the next letter (modeling phase).
Given $p$, the encoder maps the actual next letter $x$ to a codeword of length close to the optimal code length of $\log_2(1/p(x))$ bits (coding phase).
Decoding involves reversing this procedure, namely restoring $x$ given the codeword and $p$.
Arithmetic Coding is the de facto standard en-/decoder, since it closely approximates the optimal code length \cite{hbdc}.
Note that by this procedure a model \emph{assigns a code length} to an input sequence (assuming optimal encoding).
Therefore, the model essentially determines the statistical compression algorithm, since it is common to encoding and decoding.

All of the aforementioned statistical data compression algorithms base their predictions on context-conditioned Elementary Models (EMs).
EMs are typically based on simple closed form expressions, such as relative letter frequencies.
As these model contain many such EMs, they have a great impact both, on the empirical performance and associated theoretical guarantees of a statistical compressor \cite{mattern16,cts12,ptw13}.
Hence, it is wise to carefully study theoretical guarantees and empirical properties of EMs.
Theoretical guarantees on a model are typically expressed in terms of a \emph{code length analysis}.
For a code length analysis of some model \wrt\ a competitor model we typically bound the code length a model assigns to an input sequence from above by the code length assigned by the competitor model plus the excess code length which is known in the literature as the redundancy.
Here, we use a Piecewise Stationary Source (PWS) as the competitor model.
A PWS divides an input sequence into segments and predicts a fixed distribution for each segment.

In this work we focus on a generalized version of Probability Smoothing (PS) introduced in \cite{mattern11}, and provide a new code length analysis.
PS is the main EM employed by the state of the art PAQ family of statistical data compression algorithms.
We also confirm anecdotal evidence that PS surpasses the performance of various other popular EMs with similar computational complexity and hence that it is one of the main drivers of PAQ's practical success \cite{mattern16}.

\paragraph{Previous Work.}
Major approaches to EMs can roughly be divided into two groups, namely \emph{frequency-based EMs} and \emph{probability-based EMs}.
Approaches not fitting in these two categories are Finite State Machines \cite{ingber06,meron04,meron04isw,rajwan00} and Weighted Transition Diagrams \cite{shamir99,ptw13,willems96,willems97}.
Discussing these in greater detail is beyond the scope of this paper, for a survey see \cite{mattern16}.
In the following redundancy ``$O(S\cdot \dots)$'' is \wrt\ PWSs with $S$ segments, redundancy ``$O(\dots)$'' (no dependency on $S$) is \wrt\ a Stationary Source (PWS with $S=1$) and $T$ is the sequence length.

Frequency-based EMs maintain approximate letter frequencies online and predict by forming relative frequencies.
Typically these approximate letter frequencies are recency-weighted to give more emphasis to recent observations.
In practice this often improves compression.
Recency-weighting can be achieved by maintaining frequencies over a sliding window \cite{meron04isw,rashid05,ryabko97}, by resetting frequencies \cite{shamir99} or by scaling down frequencies at regular intervals \cite{howard91,mattern15rfd}.
The redundancy of these approaches typically ranges from $O(S\cdot\sqrt{T\log T})$ to $O(S\cdot\sqrt T \log T)$.

Probability-based EMs work by directly maintaining a distribution online.
A popular example is PS which was used by most members of the PAQ family of statistical data compression algorithms.
Given a PS prediction $p$ (from the previous step or from initialization) and a new letter $y$ PS first shrinks the probability of any letter $x$ to $\alpha\cdot p(x)$ and then increases the probability of $y$ by $1-\alpha$.
For a binary alphabet this approach has redundancy $O(\sqrt T)$ \cite{mattern15esp}.
An extension is to introduce a (probability) share factor parameter $\eps$ \cite{mattern11}:
After probability shrinking the probability of $y$ is increased by $(1-\alpha)\cdot(1-\eps)$ and the probability of all other observations is uniformly increased by $(1-\alpha)\cdot\frac{\eps}{N-1}$ (\cite{mattern11} only considered $N=2$).
Experiments on real-world data indicate that setting $\eps>0$ improves compression \cite{mattern11}.
Note that this approach is related to share-based expert tracking \cite{herbster98, orseau17} and to observation uncertainty \cite{mattern11}.
Another approach is to apply online convex programming to estimate distributions, for example a method based on Online Mirror Descent leads to redundancy $O(\log T)$ \cite{raginsky09}.

\paragraph{Our Contribution.}
In this work we present a novel code length analysis of the PS variant proposed in \cite{mattern11} \wrt\ PWS.
This analysis improves over previous results \cite{mattern15esp,mattern16}:
First, it applies to a more general PS variant by allowing for a non-zero share factor;
second, it remains valid for a non-binary alphabet;
third, it characterizes the redundancy \wrt\ PWSs not by the number of segments, but by the variation of PWS distributions across segments, which leads to much tighter bounds for PWS that drift slowly.
Further, experiments indicate that among other methods that take time $O(N)$ per step PS outperforms for highly non-stationary data and is close to the higher complexity method PTW.

The remaining part of this work is structured as follows:
In \secref{sec:notation} we introduce our notation and in \secref{sec:the-models} we formally define PS and PWSs.
We provide the code length analysis in \secref{sec:analysis} and experimental results in \secref{sec:experiments}.
Finally, \secref{sec:conclusion} ends this paper with a conclusion.


\section{Notation}
\label{sec:notation}

\paragraph{Sequence, Alphabet and Family.}
For integers $i$ and $j$ we use $x_{i:j}$ to denote a \emph{sequence} $x_i x_{i+1} \dots x_j$ of objects (numbers, letters, \dots).
Let $\X := \{1, 2, \dots, N\}$ denote a \emph{finite alphabet} of cardinality $N\geq2$.
Unless stated explicitly to the contrary sequences refer to sequences over $\X$.
If $i>j$, then $x_{i:j} := \phi$, where $\phi$ is the empty sequence;
if $j=\infty$, then $x_{i:j} = x_i x_{i+1} \dots$ has infinite length
and we abbreviate $x_{<i} := x_{1:i-1}$.
For objects $x_j, x_k, \dots$ with labels from $I=\{j, k, \dots\}$ we call the indexed multiset $\{x_i\}_{i\in I}$ a \emph{family}.

\paragraph{Interval, Partition and Transition Set.}
Let $[i, j) := \{i, i+1, \dots, j-1\}$ denote an interval of \emph{integers}.
A \emph{partition} of $[i,j)$ is a set $\{[t_1, t_2), [t_2, t_3), \dots, [t_n, t_{n+1})\}$ of intervals \st\ $i=t_1<t_2< \dots< t_{n+1}=j$;
the $k$-th segment is $[t_k, t_{k+1})$.
The \emph{transition set} of a partition $\P$ is defined as $\{(t,A,B)\mid A = [i, t)\text{ and } B=[t, j)\text{, for } A,B\in\P\}$, for instance, the transition set of $\{[1, 3), [3, 8), [8, 9)\}$ is $\{(3, [1, 3), [3, 8)), (8, [3, 8), [8, 9))\}$.

\paragraph{Distribution and Model.}
In general, all distributions are over $\X$.
The \emph{variation in probability mass} of two distributions $p$ and $q$ is $\lVert p-q\rVert := \sum_{x\in\X} \lvert p(x)-q(x)\rvert$.
A \emph{model} \mdl\ maps a sequence $x_{1:t}$ to a distribution $p$, its \emph{prediction}.
We define the shorthands $\mdl(x_{1:t}) := p$ and $\mdl(x; x_{1:t}) := p(x)$.
(Typicall $p$ varies with $x_{1:t}$.)

\paragraph{Codelength, Entropy and KL Divergence.}
To avoid a cluttered notation code length is measured in \textbf{nats}, thus $\log :=\log_e$.
For distributions $p$ and $q$, where $q(x)>0$, for all $x\in\X$, let $D(p\parallel q) := \sum_{x: p(x)>0} p(x)\log(p(x)/q(x))$ denote their KL divergene and let $H(p) := \sum_{x: p(x)>0} p(x)\log(1/p(x))$ denote the entropy of $p$.
A model \mdl\ \emph{assigns code length} $\ell_{\mdl}(x_{1:T}) := \sum_{1\leq t\leq T} \log(1/\mdl(x_t; x_{<t}))$ to the sequence $x_{1:T}$.


\section{The Models}
\label{sec:the-models}

\paragraph{Probability Smoothing.}
First let us formally define our model of interest.
\begin{dfn}
\label{dfn:ps}
A \emph{Probability Smoothing (PS)} model \ps\ is induced by a sequence $\alpha_{1:\infty}$ of smoothing rates, satisfying $0<\alpha_1, \alpha_2, \dots<1$, a sequence of share factors $\eps_{1:\infty}$, where $0\leq\eps_1, \eps_2, \dots\leq 1-\frac1N$ and an initial distribution $p$ \st\ $p(x)\geq\frac{\eps_1}{N-1}$, for all $x\in\X$.
The model \ps\ predicts
\(
  \ps(x; x_{1:t}) :=
  \begin{cases}    
    \alpha_t\cdot\ps(x; x_{<t}) + (1-\alpha_t)\cdot (1-\eps_t)\text,&\text{if $x=x_t$ and $t>0$,}\\
    \alpha_t\cdot\ps(x; x_{<t}) + (1-\alpha_t)\cdot \frac{\eps_t}{N-1}\text,&\text{if $x\neq x_t$ and $t>0$,}\\
    p,&\text{if $t=0$.}
  \end{cases}
\)
\end{dfn}
In the upcoming analysis we will focus on the way PS adjust its predictions over time.
To ease the presentations we will refer to the adjustment of a distribution estimate $p$ given a new letter $y$ by
\(
  \update_{\alpha,\eps}: p, y \mapsto p'
  \quad\text{\st}\quad
  p'(x) =
    \begin{cases}
      \alpha\cdot p(x) + (1-\alpha)\cdot(1-\eps)\text,& \text{if $x=y$,}\\
      \alpha\cdot p(x) + (1-\alpha)\cdot\frac \eps{N-1}\text,& \text{if $x\neq y$.}
    \end{cases}
  \eqlabel{eq:ps-update}
\)
\paragraph{Piecewise Stationary Sources.}
We now proceed by formally defining the competitor model.
\begin{dfn}
\label{dfn:pws}
A \emph{Piecewise Stationary Source (PWS)} model \pws\ for sequences of length $T$ is induced by a partition $\P$ of $\{1, 2, \dots, T\}$ and a family $\{p_S\}_{S\in\P}$ of distributions.
The model predicts $\pws(x_{<t}) := p_S$, where $S\in\P$ is the unique segment containing $t$.
\end{dfn}
The PWS code length assignment to a sequence works as follows:
For a model \pws\ with partition $\P$ every segment $S\in\P$ has an associated distribution $p_S$.
Within a segment $S$, \ie\ for $t\in S$, a letter $x_t$ receives code length $\log(1/p_S(x_t))$.
Summing yields
\(
  \ell_{\pws}(x_{1:T}) = \sum_{\substack{t\in S,\\S\in\P}} \log\frac1{p_S(x_t)}
  \text.
\)
Clearly, a PWS may have a varying degree of sophistication.
Intuitively, the more segments its partition has, the more complex it is.
Also, when distributions associated to segments greatly vary from one segment to the other then the complexity seems high.
(These two effects may also overlap.)
We formalize this intuition as follows:
\begin{dfn}
\label{dfn:pws-complexity}
Let \pws\ be a PWS model with parameters $(\P, \{p_s\}_{s\in\P})$ and let $\T$ be the transition set of $\P$.
The complexity of \pws\ is
\(
  \C_{\pws} := 1+\sum_{\mathclap{(t,A,B)\in\T}}\mkern4mu \lVert p_B-p_A\rVert
  \text.
\)
\end{dfn}
We will later see that the higher the competing PWS' complexity, the harder it is for PS to do well (the higher redundancy is).


\section{Analysis}
\label{sec:analysis}

\paragraph{Outline.}
We base our code length analysis on the well-known progress invariant technique \cite{helmbold99,zinkevich03}.
Let us now sktech this approach informally to make it more clear:
Suppose that in step $t$ PS predicts $p=\ps(x_{<t})$, we observe the letter $x_t=y$ and the PS update yields $p'=\ps(x_{1:t}) = \update_{\alpha_t,\eps_t}(p, y)$.
In this setting we bound the redundancy $\log(1/p(y))-\log(1/q(y))$ of coding $y$ from above \wrt\ an arbitrary distribution $q$.
The upper bound depends on the \emph{progress} PS makes towards $q$:
Before observing $y$ the proximity of the PS prediction $p$ and $q$ is $D(q\parallel p)$, after observing $y$ it is $D(q\parallel p')$, hence, we made progress $D(q\parallel p)-D(q\parallel p')$.
By applying this argument over multiple steps (\ie\ summing) the \textbf{per-step} progress towards some $q$ that actually reflects the inputs statistics should decrease, since the PS predictions reflect the inputs statistics more and more (\ie\ the total progress is big).
The accumulated progress will be our redundancy estimate.
Throughout the analysis we will make the following assumpions:
\begin{asm}
\label{asm:ps-fixed-parameters}
\renewcommand\theenum{(\alph{enum})}
We consider PS models with parameters $(\alpha_{1:\infty}, \eps_{1:\infty}, p)$ \st\
\begin{enum}
\item
  the smoothing rate is constant, $\alpha_1=\alpha_2=\dots=\alpha$, for $0<\alpha<1$,
\item\label{it:asm-eps}
  the share factor is constant $\eps_1=\eps_2=\dots=\eps$, for $0\leq\eps\leq1-\frac1N$, and
\item\label{it:asm-p}
  the initial estimate satisfies $p(x)\geq\frac\eps{N-1}$, for all $x\in\X$.
\end{enum}
As a shorthand we say that the PS model has (fixed) parameters $(\alpha,\eps,p)$.
\end{asm}
Note that by \asmref{asm:ps-fixed-parameters} (and \dfnref{dfn:ps}) the PS predictions satisfy
\(
  \frac\eps N\leq \ps(x; x_{1:t}) \leq 1-\eps
  \text{, for any sequence $x_{1:t}$.}
  \eqlabel{eq:ps-min-probability}
\)

\paragraph{Progress Invariant.}
To prove the progress invariant we require various technical statements.
First, we make use of the inequalities
\vspace{\abovedisplayshortskip}\vspace{-\parskip}\par\noindent
\begin{minipage}{.5\linewidth}
\(
  \log\frac1z \leq\frac{1-z}z\text{, for $0<z\leq1$,}
  \eqlabel{eq:log-inequality-1}
\)
\end{minipage}
\begin{minipage}{.5\linewidth}
\(
  \frac1z\log(1+z) \geq\frac1{1+z}\text{, for $z>0$,}
  \eqlabel{eq:log-inequality-2}
\)
\end{minipage}
\vspace{\belowdisplayshortskip}\vspace{-\parskip}\par\noindent
which may be verified by simple calculus, hence we omit their proof.
Second, we require the following:
\begin{lem}[$\boldsymbol\eps$-Proximity]
\label{lem:eps-proximity}
For arbitrary distributions $p$ and $q$ and some letter $y$ the distributions $p_0' = \update_{\alpha,0}(p, y)$ and $p_\eps' = \update_{\alpha,\eps}(p, y)$ satisfy
\(
  D(q\parallel p_0') \geq D(q\parallel p_\eps') - \log\frac1{1-\eps}
  \text.
\)
\end{lem}
\begin{prf}
Let us first consider the bound
\(
  D(q\parallel p_{\eps}')-D(q\parallel p_0')
    &=
      q(y)\log\frac{p_0'(y)}{p_\eps'(y)} + 
      \sum_{x\neq y} q(x)\log\frac{p_0'(x)}{p_\eps'(x)}\\
    &\stackrel{\text{\ref{it:eps-proximity-step1}}}\leq
      \log\frac{p_0'(y)}{p_\eps'(y)}
    \stackrel{\text{\ref{it:eps-proximity-step2}}}=
      \log\frac{p_0'(y)}{p_0'(y)-(1-\alpha)\eps}
    \stackrel{\text{\ref{it:eps-proximity-step3}}}\leq
      \log\frac1{1-\eps}
      \text,
      \eqlabel{eq:eps-proximity-1}
\)
where we have used
\renewcommand\theenum{(\alph{enum})}
\begin{enum}
\item\label{it:eps-proximity-step1}
  $q(y)\leq1$ and $p_0'(x)\leq p_\eps'(x)$, for $x\neq y$ (by \dfnref{dfn:ps}), 
\item\label{it:eps-proximity-step2}
  $p_\eps(y) = p_0(y) -(1-\alpha)\eps$ (by \dfnref{dfn:ps}) and
\item\label{it:eps-proximity-step3}
  ${p_0'(y)}/({p_0'(y)-(1-\alpha)\eps})$ is decreasing in $p_0'(y)$, hence maximal for minimum $p_0'(y)$, that is $p_0'(y) = \alpha p_0(y) + 1-\alpha \geq1-\alpha$.
\end{enum}
Rearranging \eqref{eq:eps-proximity-1} concludes the proof.
\hfill\qed
\end{prf}

\begin{lem}[Progress Invariant]
\label{lem:progress-invariant}
For arbitrary distributions $p,q$, the uniform distribution $u$ and  a letter $y$ the distribution $p' = \update_{\alpha,\eps}(p,y)$ satisfies
\(
  \log\frac1{p(y)}-\log\frac1{q(y)}
    \leq
      \frac1{\log \frac1\alpha}&\cdot\left[D(q\parallel p)-D(q\parallel p') + \log\frac1{1-\eps}\right] + \\
      \quad N&\cdot\left[D(u\parallel p)-D(u\parallel p') + \log\frac1{(1-\eps)\alpha}\right]
      \text.
      \eqlabel{eq:progress-invariant}
\)
\end{lem}
\begin{prf}
For the proof we distinguish two cases:
\begin{enumcases}
\item{$\eps=0$.}
For this case we first simplify the two KL difference terms on the \rhs\ of \eqref{eq:progress-invariant}.
Based on the simplifications we reduce \eqref{eq:progress-invariant} to an equivalent problem, that is, proving an upper bound on a function of $q$.
Finally we maximize this function to prove the upper bound.

\block{Simplifying KL differences.}
For an arbitrary distribution $r$ we obtain
\(
D(r\parallel p) - D(r\parallel p')
  = \sum_{x\in\X} r(x)\log \frac{p'(x)}{p(x)}
  &\stackrel{\eqref{eq:ps-update}}= r(y) \log\frac{p'(y)}{p(y)} - (1-r(y)) \log \frac1{\alpha}\\
  &= r(y) \log\frac{p'(y)}{\alpha p(y)} - \log\frac1{\alpha}
    \text,
    \eqlabel{eq:progress-invariant-kl}
\)
where we used $p'(x) = \alpha p(x)$, for $x\neq y$, by \eqref{eq:ps-update}.

\block{Problem Reduction and Simplified Notation.}
With a slight abuse of notation let $p:=p(y)$, $p':=p'(y)$ and $q:=q(y)$.
(Note that these now denote probabilities, \textbf{not distributions}!)
By substituting \eqref{eq:progress-invariant-kl} for either KL divergence difference term into \eqref{eq:progress-invariant}, then applying the just introduced notation and rearranging we get (recall $\eps=0$)
\(
\underbrace{\log\frac1p - \log\frac1q - \frac1{\log\frac1\alpha}\cdot\left[q\log\frac{p'}{\alpha p} - \log\frac1{\alpha}\right]}_{=:f(q)}
  \leq \underbrace{\log\frac{p'}{\alpha p}}_{=: F}
    \text,
    \eqlabel{eq:progress-invariant-reduction}    
\)
which is equivalent to \eqref{eq:progress-invariant}.

\block{Maximization.}
We now determine the maximizer $q^*$ of $f(q)$ and proceed by showing $f(q^*)\leq F$ to prove \eqref{eq:progress-invariant-reduction}.
By examining the first and second derivative of $f$,
\(
f'(q) = \frac1q-\frac{\log\frac{p'}{\alpha q}}{\log\frac1\alpha}
\quad\text{and}\quad
f''(q) = -\frac1{q^2}\text,
\)
we see that $f$ is concave and maximized for $q^*=\log\frac1\alpha/\log\frac{p'}{\alpha p}$.
Hence,
\(
f(q^*) 
  =
    \log\frac1{\frac p{\log\frac1\alpha}\cdot\log\frac{p'}{\alpha p}}
  \stackrel{\eqref{eq:ps-update},\, \eqref{eq:log-inequality-1}}\leq
    \log\frac1{p \cdot\frac{\alpha}{1-\alpha}\cdot\log\left(1+\frac{1-\alpha}{\alpha p}\right)}
  &\stackrel{\eqref{eq:log-inequality-2}}\leq
    \log\left(1+\frac{1-\alpha}{\alpha p}\right)
  \stackrel{\eqref{eq:ps-update},\, \eqref{eq:progress-invariant-reduction}}= F
    \text,
\)
where we used \eqref{eq:log-inequality-1} with $z=\alpha$, \eqref{eq:log-inequality-2} with $z=\frac{1-\alpha}{\alpha p}$ and we substituted $\frac{p'}{\alpha p} = 1+\frac{1-\alpha}{\alpha p}$, by \eqref{eq:ps-update}.
\item{$\eps>0$.}
Let us write $p'=p_\eps'$ to emphasize the dependence of $p'$ on $\eps$.
By Case~1 we have
\(
\log\frac1{p(y)}-\log\frac1{q(y)}
  &\leq
    \frac{D(q\parallel p)-D(q\parallel p_0')}{\log \frac1\alpha} + N\cdot\left[\log\frac1{\alpha} + D(u\parallel p)-D(u\parallel p_0')\right]
    \text.
\)
It remains to bound the KL terms $D(q\parallel p_0')$ and $D(u\parallel p_0')$ from below using \lemref{lem:eps-proximity} and to rearrange to conclude the proof.
\hfill\qed
\end{enumcases}
\end{prf}

\paragraph{Putting it all Together.}
We are now almost ready to carry out the code length analysis.
First we need some technical lemmas.
\begin{lem}[KL Difference L1-Bound]
\label{lem:kl-difference-l1-bound}
For distributions $p$ and $w$, both with probability at least $m>0$ on any letter, and any distribution $v$ we have
\(
  D(w\parallel p)-D(v\parallel p) \leq \log\frac1m \cdot \left\lVert w-v\right\rVert
  \text.
\)
\end{lem}
\begin{prf}
For the proof we establish a Taylor expansion of $D(v\parallel p)$ and rearrange.
\block{Taylor expansion.}
Since $\partial D(v\parallel p)/\partial v(x) = 1+\log(v(x)/p(x))$ and since $D(v\parallel p)$ is convex in $v$ we have
\(
  D(v\parallel p)
  &\geq
    D(w\parallel p) + \sum_{x\in\X} \left(1+\log\frac{w(x)}{p(x)}\right) \cdot (v(x)-w(x))\\
  &=
    D(w\parallel p) + \sum_{x\in\X} \log\frac{w(x)}{p(x)} \cdot (v(x)-w(x))
    \eqlabel{eq:kl-difference-l1-bound-0}
    \text.
\)
\block{Rearranging.}
By the above we get
\(
  D(w\parallel p)-D(v\parallel p)
  &\stackrel{\eqref{eq:kl-difference-l1-bound-0}}\leq
    \sum_{x\in\X} \log\frac{w(x)}{p(x)} \cdot (w(x)-v(x))
  \leq
    \sum_{x\in\X} \left\lvert\log\frac{w(x)}{p(x)}\right\rvert \cdot \lvert v(x)-w(x)\rvert
    \\
  &\leq
    \max_{x\in\X} \left\lvert\log\frac{w(x)}{p(x)}\right\rvert \cdot \sum_{x\in\X} \lvert v(x)-w(x)\rvert
  \leq
    \log\frac1m \cdot \lVert v-w\rVert
    \text,
\)
where for the last inequality we used $m\leq \frac{w(x)}{p(x)}\leq\frac1m$.
\hfill\qed
\end{prf}
We are now ready to state and prove the main result of this work.
\begin{thm}[Main Result]
\label{thm:ps-fixed-parameters}
Let \pws\ be a PWS model for sequences of length $T$ and let $u$ be the uniform distribution.
The PS model \ps\ with parameters $(\alpha,\eps,p)$ guarantees
\(
  \ell_{\ps}(x_{1:T})
  \leq
    \ell_{\pws}(x_{1:T})
    &+ \left[\frac{\log\frac1{1-\eps}}{\log\frac1\alpha} + N\log\frac1{\alpha} + (N+1)\log\frac1{1-\eps}\right]\cdot T\\
    &+ \phantom{\Bigg[} \frac{\log{\frac N\eps}}{\log\frac1\alpha}\cdot \C_{\pws} + N \cdot D(u\parallel p)
    \text.    
\)
\end{thm}
\begin{prf}
For the proof we first introduce $\pws'$, a slight modification of $\pws$.\footnote{This is necessary, since we use \lemref{lem:kl-difference-l1-bound} which relies on probabilities bounded away from $0$.}
We then bound the redundancy of \ps\ \wrt\ $\pws'$, simplify araising telescoping KL difference terms and finally remove the dependency on $\pws'$ \st\ the redundancy bound is \wrt\ \pws, as desired.
\block{Definition and Properties of $\pws'$.}
Let $(\P, \{u_S\}_{S\in\P})$ be the parameters of \pws\ and let $\pws'$ have parameters $(\P, \{v_S\}_{S\in\P})$, where
\(
  v_S(x) = (1-\eps)\cdot u_S(x) + \frac\eps N
  \text.
  \eqlabel{eq:ps-fixed-smoothing-rate-0}
\)
Note that $\pws'$ has two useful properties that we will exploit later.
First, given two segments $A$ and $B$ from $\P$ we have $v_B(x)-v_A(x) = (1-\eps)\cdot(u_B(x)-u_A(x))$ (by \eqref{eq:ps-fixed-smoothing-rate-0}), which implies $\lVert v_B-v_A\rVert = (1-\eps)\cdot\lVert u_B-u_A\rVert$, and combined with \dfnref{dfn:pws-complexity} this leads to
\(
  \C_{\pws'}\leq \C_{\pws}
  \text.
  \eqlabel{eq:ps-fixed-smoothing-rate-1a}
\)
Second, for some segment $S$ from $\P$ we have $v_S(x) \geq (1-\eps)\cdot u_S(x)$ (by \eqref{eq:ps-fixed-smoothing-rate-0}) implying $\log(1/v_S(x)) \leq \log(1/u_S(x)) + \log(1/(1-\eps))$, so
\(
  \ell_{\pws'}(x_{1:T})\leq \ell_{\pws}(x_{1:T}) + \log\frac1{1-\eps} \cdot T
  \text.
  \eqlabel{eq:ps-fixed-smoothing-rate-1b}
\)
\block{Competing with $\pws'$.}
For brevity let $p_t := \ps(x_{<t})$ and $d_t(q) = D(q\parallel p_t)$ and consider some segment $S$ from $\P$ and let $t\in S$.
Given this, \lemref{lem:progress-invariant} yields
\(
  \log\frac1{p_t(x_t)}-\log\frac1{v_S(x_t)}
  \stackrel{\nref[L.\,]{lem:progress-invariant}}\leq
    \frac1{\log\frac1\alpha}&\cdot\left[ d_t(v_S)-d_{t+1}(v_S)+\log\frac1{1-\eps}\right] + \\
    N&\cdot\left[d_t(u)-d_{t+1}(u)+\log\frac1{\alpha(1-\eps)}\right]
    \text,
    \eqlabel{eq:ps-fixed-smoothing-rate-1}
\)
where $u$ is the uniform distribution.
By summing \eqref{eq:ps-fixed-smoothing-rate-1} over all steps $t\in S$, for every segment $S\in\P$, we get
\(
  \ell_{\ps}(x_{1:T})-\ell_{\pws'}(x_{1:T})
  \stackrel{\eqref{eq:ps-fixed-smoothing-rate-1}}\leq
    \frac1{\log\frac1\alpha}\cdot\sum_{\substack{t\in S,\\S\in\P}} (d_t(v_S)-d_t(v_S)) +
    N\cdot\sum_{\mathclap{1\leq t\leq T}} (d_t(u)-d_{t+1}(u)) \\
  \omit{\hfill$\displaystyle
    +\mkern4mu\left[\frac{\log\frac1{1-\eps}}{\log\frac1\alpha} + N\log\frac1{\alpha(1-\eps)}\right]\cdot T
    \text.$}\quad
    \eqlabel{eq:ps-fixed-smoothing-rate-2}
\)
\block{Simplifying Telescoping Sums.}
Let $F$ be the first segment from $\P$ and let $L$ be the last segment from $\P$.
By simplifying the sums in \eqref{eq:ps-fixed-smoothing-rate-2} we get
\(
  \sum_{\substack{t\in S,\\S\in\P}} (d_t(u)-d_{t+1}(u))
    &= \sum_{1\leq t\leq T}(d_t(u)-d_{t+1}(u))
    \stackrel{\text{\ref{it:ps-fixed-smoothing-rate-step0}}}= d_1(u)-d_{T+1}(u)
    \stackrel{\text{\ref{it:ps-fixed-smoothing-rate-step1}}}\leq D(u\parallel p)
    \text,
  \eqlabel{eq:ps-fixed-smoothing-rate-3}
  \\
  \sum_{\substack{t\in S,\\S\in\P}} (d_t(v_S)-d_{t+1}(v_S))
    &= \sum_{S\in\P} \sum_{t\in S} (d_t(v_S)-d_{t+1}(v_S))
    \stackrel{\text{\ref{it:ps-fixed-smoothing-rate-step0}}}= \mkern8mu\sum_{\mathclap{S=[a,b)\in\P}}\mkern6mu(d_a(v_S)-d_b(v_S))\\
    &\stackrel{\text{\ref{it:ps-fixed-smoothing-rate-step2}}}= d_1(v_F)-d_{T+1}(v_L) + \mkern2mu \sum_{\mathclap{(t,A,B)\in\T}}\mkern4mu (d_t(v_B)-d_t(v_A))\\
    &\stackrel{\text{\ref{it:ps-fixed-smoothing-rate-step3}}}\leq \log\frac N\eps + \mkern2mu \sum_{\mathclap{(t,A,B)\in\T}}\mkern4mu (d_t(v_B)-d_t(v_A))\\
    &\stackrel{\text{\ref{it:ps-fixed-smoothing-rate-step4}}}\leq \log\frac N\eps + \log\frac N\eps\mkern2mu\cdot\mkern2mu\sum_{\mathclap{(t,A,B)\in\T}}\mkern4mu \lVert v_B-v_A\rVert
    \stackrel{\nref[D.\,]{dfn:pws-complexity}}= \log\frac N\eps\cdot \C_{\pws'}
    \text,
  \eqlabel{eq:ps-fixed-smoothing-rate-4}
\)
where we have used:
\renewcommand\theenum{(\alph{enum})}
\begin{enum}
\item\label{it:ps-fixed-smoothing-rate-step0}
  The sum telescopes.
\item\label{it:ps-fixed-smoothing-rate-step1}
  We have $d_1(u) = D(u\parallel p_1) = D(u\parallel p)$ and $d_{T+1}(u) = D(u\parallel p_{T+1})\geq0$.
\item\label{it:ps-fixed-smoothing-rate-step2}
  Suppose that $\P$ is \st\ its segments, in order, are $F=[1,i)$, $G=[i, j)$, \dots, $L=[k, T+1)$.
  We have
  \(
    \smash{\sum_{\mathclap{S=[a,b)\in\P}}}\mkern4mu (d_a(v_S)-d_b(v_S))
      &= d_1(v_F)-d_i(v_F) + d_i(v_G)-d_j(v_G) + \dots\\
      &= d_1(v_F) + (d_i(v_G)-d_i(v_F)) + \dots + (\dots) - d_{T+1}(v_L)\\
      &= d_1(v_F) - d_{T+1}(v_L) + \sum_{\mathclap{(t,A,B)\in\T}} \mkern4mu (d_t(v_B)-d_t(v_A))
      \text.
  \)
\item\label{it:ps-fixed-smoothing-rate-step3}
  We have $d_1(v_F) = D(v_F\parallel p_1) = D(v_F \parallel p) \leq \max_{x\in\X} \log\frac1{p(x)} \leq \log\frac N\eps$, by \eqref{eq:ps-min-probability}, and $d_{T+1}(v_L) = D(v_L\parallel p_{T+1})\geq0$.
\item\label{it:ps-fixed-smoothing-rate-step4}
  \lemref{lem:kl-difference-l1-bound} applied to $d_t(v_B)-d_t(v_A) = D(v_B\parallel p_t)-D(v_A\parallel p_t)$ with $m=\frac \eps N$, since $v_B(x)\geq m$ (by \eqref{eq:ps-fixed-smoothing-rate-0}) and $p_t(x)\geq m$, by \eqref{eq:ps-min-probability}, both for all $x\in\X$.
\end{enum}
\block{Competing with $\pws$.} 
By plugging \eqref{eq:ps-fixed-smoothing-rate-3} and \eqref{eq:ps-fixed-smoothing-rate-4} into \eqref{eq:ps-fixed-smoothing-rate-2} we get
\(
  \ell_{\ps}(x_{1:T})-\ell_{\pws'}(x_{1:T})
  \stackrel{\mathclap{\eqref{eq:ps-fixed-smoothing-rate-2}\,\text-\,\eqref{eq:ps-fixed-smoothing-rate-4}}}\leq
    \frac1{\log\frac1\alpha}\cdot\log\frac N\eps\cdot\C_{\pws'} +
    N\cdot D(u\parallel p) \hphantom{\hspace{.15\linewidth}}\\
  \omit{\hfil$\displaystyle
    +\mkern4mu\left[\frac{\log\frac1{1-\eps}}{\log\frac1\alpha} + N\log\frac1{\alpha(1-\eps)}\right]\cdot T
    $.}
    \eqlabel{eq:ps-fixed-smoothing-rate-5}
\)
To end the proof we bound the \lhs\ of \eqref{eq:ps-fixed-smoothing-rate-5} from below by \eqref{eq:ps-fixed-smoothing-rate-1b} and the \rhs\ from above by \eqref{eq:ps-fixed-smoothing-rate-1a} and rearrange.
\hfill\qed
\end{prf}
\paragraph{Discussion.}
Let us now discuss the code length bound given in \thmref{thm:ps-fixed-parameters}.
The major contribution in redundancy is twofold:

First, regardless of the competing PWS, the redundancy will be high if either $\alpha$ is too small (PS predictions vary extremely from step to step, adaption is too fast)  or if $\alpha$ is too large (PS will barely adjust its predictions, adaption is too slow).
Furthermore, the range of possible PS predictions must be sufficiently rich (\ie\ $\eps$ shouldn't be too large) to enable adaption at all.
The following term captures this, since it penalizes too large or too small $\alpha$ and large $\eps$, proportional to $T$,
\(
  \left[\frac{\log\frac1{1-\eps}}{\log\frac1\alpha} + N\log\frac1{\alpha} + (N+1)\log\frac1{1-\eps}\right]\cdot T  
  \text.
\)

Second, if we want compete with a complex PWS that well reflects the inputs statistics, then the smoothing rate $\alpha$ (weight of old observations) must be small to be able to adapt to the input quickly.
Moreover, the probabilities PS predicts should never get too extreme (\ie\ $\eps$ shouldn't be too small), since the more extreme they are, the longer it takes to adapt to changing statistics.
These effects are captured by the following term, since it penalizes large $\alpha$ and small $\eps$, proportional to $\C_{\pws}$,
\(
  \frac{\log\frac N\eps}{\log\frac1\alpha}\cdot\C_{\pws}
  \text.
\)

Clearly, the choice of the parameters $\alpha$ and $\eps$ is a tradeoff between all those aspects.
It especially depends on the complexity of desirable PWS, which is unknown in general.
Hence, in the following we just give an example of choosing those parameters that is independent of $\C_{\pws}$.
If $T\geq2$ and if we set
\(
  \alpha=e^{-\sqrt{\log(NT)/(NT)}}
  \quad\text{and}\quad
  \eps=\frac1T
  \text,
  \eqlabel{ps-fixed-parameters}
\)
then \thmref{thm:ps-fixed-parameters} states that  we have
\(
  \ell_{\ps}(x_{1:T})
  \leq
    \ell_{\pws}(x_{1:T}) + (1+\C_{\pws}) &\cdot \sqrt{NT\log(NT)} \\
    \displaystyle+\mkern4mu(1+o(1))&\cdot \sqrt{NT/\log(NT)} +  O(N)
    \text.
  \eqlabel{eq:ps-redundancy-for-fixed-parameters}
\)
Assuming a fixed alphabet size $N$ the redundancy is $O(\C_{\pws}\cdot \sqrt{T\log T})$, hence sublinear, as long as the complexity satisfies $\C_{\pws} = o(\sqrt{T/{\log T}})$.

\paragraph{Extensions.}
By the above discussion it becomes clear that a ``good'' choice of fixed PS parameters depends on the sequence length $T$.
In general this quantity is unknown in advance.
However, we may lift this limitation by varying parameters with time:
We can employ a gradually increasing smoothing rate and at the same time decrease the share factor \st\
\(
  \alpha_t \approx  e^{-\sqrt{\log(Nt)/(Nt)}}
  \quad\text{and}\quad
  \eps_t \approx t^{-1}
  \text.
  \eqlabel{eq:ps-varying-parameters}
\)
This yields guarantees similar to \eqref{eq:ps-redundancy-for-fixed-parameters}.
\iffullpaper
See the appendix for details.
\fi
Another way to tackle this problem is the doubling trick.


\section{Experiments}
\label{sec:experiments}

\begin{figure}[t]
\centering
\fbox{\includegraphics{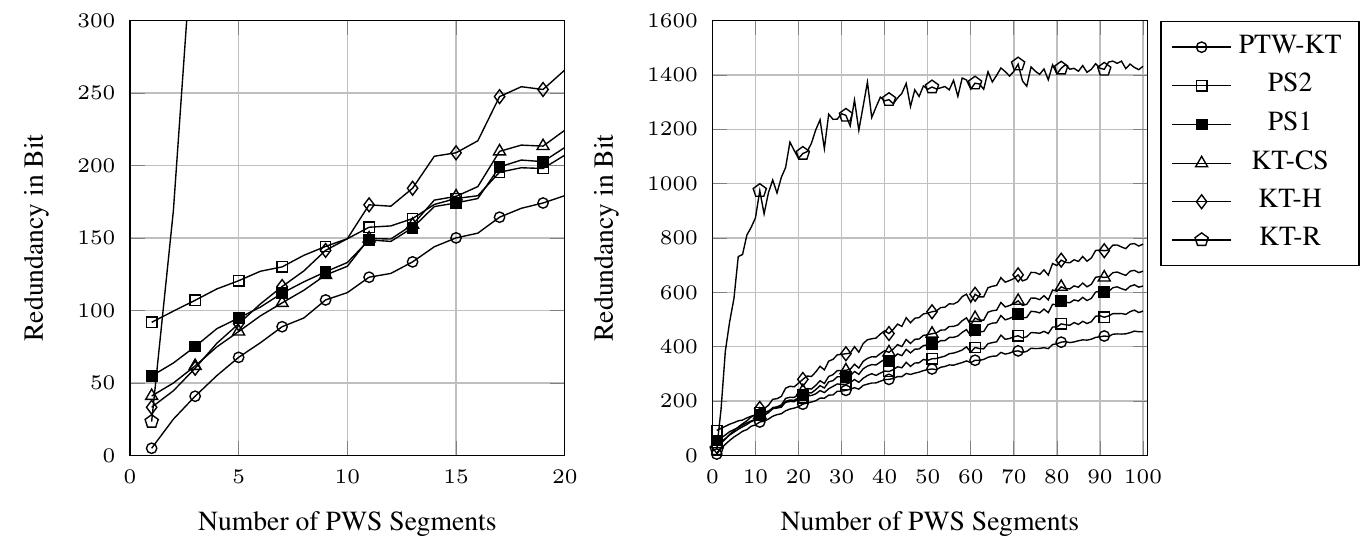}\hskip.5cm}
\caption{Average redundancy (lower is better) of various models \wrt\ PWSs with a varying number of segments for sequence length $8192$. Zoomed in (left), full view (right).}
\label{fig:pws-redundancy}
\end{figure}

\paragraph{Experimental Setup.}
In our experiments we compare the redundancy of various practical models \wrt\ PWSs on artificial data.
We consider PS with fixed parameters (PS1, see \eqref{ps-fixed-parameters}) and varying parameters (PS2, see \eqref{eq:ps-varying-parameters}), the KT model \cite{kt81} with counts aged by a discount rate of $0.98$ in every step (KT-CS) \cite{ptw13}, with counts halved every $\sqrt T$ steps (KT-H) \cite{mattern16}, with counts reset in exponentially increasing intervals of length $1, 2, 4, \dots$ (KT-R) \cite{shamir99}.
All of those models take time $O(N)$ per step.
As a reference for more complex models we also include PTW with a KT base model (PTW-KT) \cite{ptw13}.
PTW-KT takes time $O(N\log T)$ per step.
For $S\in\{1, 2, \dots, 100\}$ (number of PWS segments) we do:
1. draw the model \pws\ uniform at random from the set of all PWSs for binary sequences of length $T=8192$ with $S$ segments,
2. draw a sequence at random according to $\pws(x_{1:T})$,
3. compute the redundancy $\ell_{\mdl}(x_{1:T}) - \ell_{\pws}(x_{1:T})$ of either model \mdl\ \wrt\ \pws\ and
4. repeat steps 1. to 3. 100 times to compute redundancy averages.

\paragraph{Results.}
\figref{fig:pws-redundancy} summarizes the experimental results.
In general KT-R has worst overall performance and PTW-KT has best overall performance (recall that PTW-KT takes time $O(N\log T)$ per step).
Considering the other models there is a phase transition at around $10$ to $15$ segments.
One should think of two regimes, slightly non-stationary data (number of PWS segments is small compared to the sequence length) and highly non-stationary data (number of segments is large compared to the sequence length).
In the former case PS2 underperforms KT-CS, KT-H and PTW-KT and PS1 is on par;
in the latter case PS1 and PS2 outperform the other models except PTW-KT, remarkably PS2 is close to PTW-KT.
For models that take time $O(N)$ per step it seems that the more gentle and frequent aging takes place, the better in the regime of highly non-stationary data, hence the ordering PS1/PS2, KT-CS, KT-H, KT-R.


\section{Conclusion}
\label{sec:conclusion}
In this work we revisited a generalization of PS, the core EM in the state of the art family of PAQ statistical data compression algorithms.
Our main contribution is a code length analysis of generalized PS \wrt\ PWSs.
In particular our results hold for a finite (not necessarily binary) alphabet and relate the redundancy to the PWS complexity, a measure more fine-grained than just the number of PWS segments.
A brief experimental study shows that for highly non-stationary data PS improves over other common methods with similar time complexity and performs only slightly worse than PTW-KT.

We believe that it is worthwhile to extend this work in terms of theoretical and practical matters.
From the theory point of view it is straight forward to extend the code length analysis to Count Smoothing \cite{mattern16}, since asymptotically it is equivalent to PS.
Another extension is to consider bounds of the form ``the PS code length is (at most) within a multiplicative factor of the PWS code length plus additive terms independent of the sequence length''.
From a practical point of view we think it is important to carry out an experimental study that covers major EMs on artificial data and real world data to guide the design of statistical data compression algorithms.

\paragraph{Acknowledgement.}
The author would like to thank, in alphabetic order, Tor Lattimore, Laurent Orseau, Joel Veness and the anonymous reviewers
\iffullpaper of the Data Compression Conference\fi\
for helpful comments and corrections to this paper.

\iffullpaper\newpage\fi

\bibliographystyle{IEEEtranS}
\let\oldthebibliography=\thebibliography
\let\endoldthebibliography=\endthebibliography
\renewenvironment{thebibliography}[1]{\begin{oldthebibliography}{#1}\setlength{\itemsep}{0pt}\footnotesize}{\end{oldthebibliography}}  
\bibliography{IEEEabrv,ref}

\iffullpaper

\newpage
\appendix

\section{Varying Smoothing Rate}

\paragraph{PS Parameters.}
When the sequence length $T$ is unknown in advance we may choose the PS parameters almost as in \eqref{eq:ps-varying-parameters}, except minor modifications that don't effect the asympotics.
(As we will see later this increases the leading redundancy term by a factor of $\sqrt2$ compared to \eqref{eq:ps-redundancy-for-fixed-parameters}.)
More precisely, we consider the following choice of parameters:

\begin{asm}
\label{asm:ps-varying-parameters}
\renewcommand\theenum{(\alph{enum})}
We consider PS models with parameters $(\alpha_{1:\infty}, \eps_{1:\infty}, p)$ \st\
\begin{enum}
\item
  the smoothing rate is $\alpha_t=\exp\left( -\sqrt{\frac{\log(N/\eps_t)}{2Nt}} \right)$,
\item
  the share factor is $\eps_t = \frac1{t+1}$ and
\item
  the initial estimate satisfies $p(x)\geq\frac{\eps_1}{N-1}$, for all $x\in\X$.
\end{enum}
\end{asm}
Note that by \asmref{asm:ps-varying-parameters} (and \dfnref{dfn:ps}) the PS predictions satisfy
\(
  \alpha_1\leq \alpha_2\leq \dots &\quad\text{and}\quad \eps_1\geq \eps_2 \geq \dots
  \eqlabel{eq:ps-varying-parameters-monotonicity}
  \text,\\
  \frac{\eps_T} N&\leq \ps(x; x_{<t})
  \text.
  \eqlabel{eq:ps-varying-parameters-pmin}
\)
We will later take advantage of these properties.

\paragraph{Some Technical Statements.}
The technical parts of the upcoming analysis consist of simplyfing telescoping sums of KL differences (similar to the proof of \thmref{thm:ps-fixed-parameters}) and sums over functions of PS parameters.
Since the PS parameters vary with time the upcoming analysis requires considerable technical work.
To ease this we parenthesize the following two lemmas:

\begin{lem}
\label{lem:erfi-sum}
For $T\geq2$ we have $\sum_{1<t\leq T}\frac1{\sqrt{t \log t}} \leq 8\sqrt{\frac T{\log T}}$.
\end{lem}
\begin{prf}
We start by bounding the sum by an integral and substitute $t=f(z)$ with $f(z)=e^{z^2}$,
\(
  \sum_{1< t\leq T} \frac1{\sqrt{t\log t}}
  \leq \int_1^T \mkern-10mu \frac{\mathrm dt}{\sqrt{t\log t}}
  = \int_{f^{-1}(1)}^{f^{-1}(T)} \mkern-10mu \frac{f'(z) \mathrm dz}{\sqrt{f(z) \log f(z)}}
  = 2\mkern-2mu\int_0^{Z} \mkern-5mu e^{z^2/2} \mathrm dz
  \text,
\)
for $Z = f^{-1}(T) = \sqrt{\log T}$.
We proceed by further bounding the latter integral,
\(
  2\int_0^{Z} e^{z^2/2} \mathrm dz  
  \stackrel{\text{\ref{it:erfi-sum-0}}}\leq 4\int_{Z/2}^{Z} e^{z^2/2} \mathrm dz
  \stackrel{\text{\ref{it:erfi-sum-1}}}\leq \frac8Z \int_{Z/2}^{Z} z e^{z^2/2} \mathrm dz
  \leq \frac{8e^{Z^2/2}}Z
  = 8\sqrt{\frac T{\log T}}
  \text,
\)
where we used
\renewcommand\theenum{(\alph{enum})}
\begin{enum}
\item\label{it:erfi-sum-0}
that $e^{z^2/2}$ is increasing and
\item\label{it:erfi-sum-1}
$e^{z^2/2} \leq 2z e^{z^2/2}/Z$, for $z\geq Z/2$.
\hfill\qed
\end{enum}
\end{prf}

\begin{lem}
\label{lem:segment-sum}
Given a partition $\P$ of $\{1, 2, \dots, T\}$ let
\vskip.5ex
\renewcommand\theenum{\upshape(\alph{enum})}
\begin{enum}
\item\label{it:segment-sum-sf-sl}
  $F$ be the first segment and $L$ be the last segment of $\P$,
\item\label{it:segment-sum-wt}
  $w_1, w_2, \ldots$ be a sequence of weights \st\ $w_1\leq w_2\leq \dots$,
\item
  $d_1, d_2, \ldots : \P \rightarrow\RR$ be a sequence of functions,
\item\label{it:segment-sum-D}
  $D$ be an upper bound on their function value, \ie\ $d_t(S)\leq D$, for all $t\in S$, given any segment $S\in\P$ and
\item
  let $\T$ be the transition set of $\P$.  
\end{enum}
\vskip.5ex
It holds that
\(
\smash{\sum_{\substack{t\in S,\\S\in\P}}} w_t(d_t(S) - d_{t+1}(S))
  \leq
    w_{T{+}1} (D-d_{T{+}1}(L)) - w_1&(D-d_1(F))
    \\[4pt]
    + \sum_{\mathclap{(t, A, B) \in \T}} w_t&(d_t(B)-d_t(A))
    \text.
\)
\end{lem}
\begin{prf}
For the proof we first bound the sum over all $t\in S$ (segment sum) for an arbitrary segment $S$ from above and then sum this bound over all segments.

\block{Segment Sum.}
For some segment $S=[a,b)$ we obtain
\(
\sum_{t\in S} w_t (d_t(S)-d_{t+1}(S))
  &=
    w_a d_a(S) - w_{b-1} d_b(S) + \sum_{\mathclap{a<t<b}}\hskip4pt d_t(S)\cdot(w_t-w_{t-1})
    \\[4pt]
  &\stackrel{\mathclap{\text{\ref{it:segment-sum-wt}, \ref{it:segment-sum-D}}}}\leq
    w_a d_a(S) - w_{b-1} d_b(S) + D\cdot \sum_{\mathclap{a<t<b}}\hskip4pt (w_t-w_{t-1})
    \\[-2pt]
  &=
    w_{b-1}(D-d_b(S))-w_a(D-d_a(S))
    \vphantom{\sum_{\mathclap{a<t<b}}}
    \\[4pt]
  &\stackrel{\mathclap{\text{\ref{it:segment-sum-wt}, \ref{it:segment-sum-D}}}}\leq
    w_b (D-d_b(S))-w_a(D-d_a(S))
    \text.
    \eqlabel{eq:segment-sum-0}
\intertext{%
\block{Sum over all Segments.}
By summing over all segments we get
}
\smash{\sum_{\substack{t\in S,\\S\in\P}}} w_t(d_t(S) - d_{t+1}(S))
  &=
    \sum_{s\in\P}\sum_{t\in S} w_t (d_t(S)-d_{t+1}(S)) \vphantom{\sum_{\substack{a\\b}}}
    \\
  &\stackrel{\eqref{eq:segment-sum-0}}\leq
    \sum_{\mathclap{\substack{S\in\P,\\S=[a,b)}}} \left(w_b (D-d_b(S))-w_a(D-d_a(S))\right)
    \eqlabel{eq:segment-sum-1}
    \\
  &\stackrel{\mathclap{\text{\ref{it:segment-sum-sf-sl}}}}=
    w_{T+1} (D-d_{T+1}(L)) - w_1(D-d_1(F)) \vphantom{\sum}\\
  & \omit\hfill$\displaystyle+ \sum_{\mathclap{(t,A,B)\in\T}} w_t(d_t(B)-d_t(A))$
    \text.
\)
For the last equality we have used the same argument as in the proof of \thmref{thm:ps-fixed-parameters}, see Item (c) in that proof.
\hfill\qed
\end{prf}

\paragraph{Code Length Analysis.}
We are now ready to analyze PS for the parameter choice given above.

\begin{thm}
Let \pws\ be a PWS model for sequences of length $T$, let $T\geq 2$ and let $u$ be the uniform distribution.
The PS model \ps\ with parameters given in \asmref{asm:ps-varying-parameters} guarantees
\(
  \ell_{\ps}(x_{1:T})
  \leq \ell_{\pws}(x_{1:T}) + \sqrt{2N(T+1) \log(N(T+1))}\cdot(1+\C_{\pws}) + \sqrt{\frac{128 N T}{\log T}}\\
  \omit{\hfill$\displaystyle+\mkern6mu O(N\log T)$}
  \text.
\)
\end{thm}

\begin{prf}
For the proof use the same strategy as in the proof of \thmref{thm:ps-fixed-parameters}:
First, we define $\pws'$, a refinement of $\pws$, and analyze the redundancy of \ps\ \wrt\ $\pws'$.
We then proceed by bounding several sums over terms that depend on the paramters of \ps\ to simplify the bound (\wrt\ $\pws'$).
Finally, we transform the competitive guarantees \wrt\ $\pws'$ to gurantees \wrt\ \pws.

\block{Definition and Properties of $\pws'$.}
Let $(\P, \{u_S\}_{S\in\P})$ be the parameters of \pws\ and let $\pws'$ have parameters $(\P, \{v_S\}_{S\in\P})$, where
\(
  v_S(x) = (1-\eps_T)\cdot u_S(x) + \frac{\eps_T}N
  \text.
  \eqlabel{eq:ps-varying-smoothing-rate-0}
\)
Similarly to the proof of \thmref{thm:ps-fixed-parameters} we will take advantage of the following properties,
\(
  \C_{\pws'}
  &\leq \C_{\pws}  
  \text,
  \eqlabel{eq:ps-varying-smoothing-rate-1a}
  \\
  \ell_{\pws'}(x_{1:T})
  &\leq \ell_{\pws}(x_{1:T}) + \log\frac1{1-\eps_T} \cdot T \leq \ell_{\pws}(x_{1:T}) + 1
  \text.
  \eqlabel{eq:ps-varying-smoothing-rate-1b}
\)
The last inequality is due to $\log(1/(1-\eps_T)) = \log(1+1/T)\leq 1/T$.

\block{Competing with $\pws'$.}
For brevity let $p_t := \ps(x_{<t})$ and consider some segment $S$ from $\P$ and let $t\in S$.
Given this, \lemref{lem:progress-invariant} yields
\(
  \log\frac1{p_t(x_t)}-\log\frac1{v_S(x_t)}
  \stackrel{\nref[L.\,]{lem:progress-invariant}}\leq
    \frac1{\log\frac1\alpha_t}&\cdot\left[ D(v_S\parallel p_t)-D(v_S\parallel p_{t+1})+\log\frac1{1-\eps_t}\right] + \\
    N&\cdot\left[D(u\parallel p_t)-D(u\parallel p_{t+1})+\log\frac1{\alpha_t(1-\eps_t)}\right]
    \text,
    \eqlabel{eq:ps-varying-smoothing-rate-1}
\)
where $u$ is the uniform distribution.
By summing \eqref{eq:ps-varying-smoothing-rate-1} over all steps $t\in S$, for every segment $S\in\P$, we get
{%
\small
\(
  \ell_{\ps}(x_{1:T})-\ell_{\pws'}(x_{1:T})
  &\stackrel{\eqref{eq:ps-varying-smoothing-rate-1}}\leq
    \sum_{\substack{t\in S,\\S\in\P}} \underbrace{\frac{D(v_S\parallel p_t)-D(v_S\parallel p_{t+1})}{\log\frac1{\alpha_t}}}_{\text{Term (i)}} +
    \sum_{1\leq t\leq T} \underbrace{\frac{\log\frac1{1-\eps_t}}{\log\frac1{\alpha_t}}}_{\text{Term (ii)}} +\\
  &\mathrel{\phantom\leq}N\mkern2mu\cdot\mkern-8mu\sum_{1\leq t\leq T}\bigg(\underbrace{D(u\parallel p_t)-D(u\parallel p_{t+1})\vphantom{\log\frac1{\alpha_t}}}_{\text{Term (iii)}} + \underbrace{\log\frac1{(1-\eps_t)}}_{\text{Term (iv)}} + \underbrace{\log\frac1{\alpha_t}}_{\text{Term (v)}}\bigg)
    \text.
    \eqlabel{eq:ps-varying-smoothing-rate-2}
\)
}
\block{Term (i).}
For the first term we obtain,
\(
  \sum_{\substack{t\in S,\\S\in\P}}\frac{D(v_S\parallel p_t)-D(v_S\parallel p_{t+1})}{\log\frac1{\alpha_t}}
  &\stackrel{\text{\ref{it:ps-varying-smoothing-rate-termi-step0}}}\leq \frac{\log\frac N{\eps_T}}{\log\frac1{\alpha_{T+1}}} + \mkern-8mu\sum_{(t,A,B)\in\T}\mkern-8mu \frac{D(v_B\parallel p_t) - D(v_A\parallel p_t)}{\log\frac1{\alpha_t}}\\
  &\stackrel{\text{\ref{it:ps-varying-smoothing-rate-termi-step1}}}\leq \frac{\log\frac N{\eps_T}}{\log\frac1{\alpha_{T+1}}} + \log\frac N{\eps_T}\mkern4mu\cdot\mkern-12mu\sum_{(t,A,B)\in\T}\mkern-8mu \frac{\lVert v_B-v_A \rVert}{\log\frac1{\alpha_t}}\\
  &\stackrel{\text{\ref{it:ps-varying-smoothing-rate-termi-step2}}}\leq \frac{\log\frac N{\eps_T}}{\log\frac1{\alpha_{T+1}}} \cdot \C_{\pws'}\\
  &\stackrel{\text{\ref{it:ps-varying-smoothing-rate-termi-step3}}}\leq \sqrt{2N(T+1) \log\frac N{\eps_T}}\cdot \C_{\pws'}\\
  &= \sqrt{2N(T+1)\log(N(T+1))}\cdot \C_{\pws'}
  \text.
  \eqlabel{eq:ps-varying-smoothing-rate-termi}
\)
The intermediate steps are as follows:
\renewcommand\theenum{(\alph{enum})}
\begin{enum}
\item\label{it:ps-varying-smoothing-rate-termi-step0}
We applied \lemref{lem:segment-sum} with $w_t = 1/\log\frac1{\alpha_t}$ (weights $w_t$ are non-decreasing, since $\alpha_t$ is non-decreasing), $d_t(S) = D(v_S\parallel p_t)$ and $D=\log\frac N{\eps_T}$ and dropped all negative terms.
(Note that $D$ is an upper bound on $d_t(S)$, for $t\leq T$, since by \eqref{eq:ps-varying-parameters-pmin} we have
$d_t(S) = D(v_S\parallel p_t) \leq \max_{x\in\X} \log\frac1{p_t(x)} \leq \log\frac N{\eps_T}$.)

\item\label{it:ps-varying-smoothing-rate-termi-step1}
We applied \lemref{lem:kl-difference-l1-bound} with $m=\log\frac N{\eps_T}$, since $p_t(x)\geq m$ (by \eqref{eq:ps-varying-parameters-pmin}) and $v_B(x)\geq m$ (by \eqref{eq:ps-varying-smoothing-rate-0}), both for any $x\in\X$.

\item\label{it:ps-varying-smoothing-rate-termi-step2}
We plugged in $\log\frac1{\alpha_t}\geq \log\frac1{\alpha_{T+1}}$ and used \dfnref{dfn:pws-complexity}.

\item\label{it:ps-varying-smoothing-rate-termi-step3}
We bounded $\log\frac1{\alpha_{T+1}}= \sqrt{{\log(N/\eps_{T+1})}/(2N(T+1))}\geq \sqrt{{\log(N/\eps_{T})}/(2N(T+1))}$.

\end{enum}

\block{Term (ii).}
We have $\log\frac1{1-\eps_t}=\log\frac{t+1}t\leq\frac1t$ and $\log\frac N{\eps_t}=\log(N(t+1))$.
Hence, we may simplify
\(
  \frac{\log\frac1{1-\eps_t}}{\log\frac1\alpha_t}
  = \frac{\log\frac1{1-\eps_t}}{\sqrt{\log\frac N{\eps_t}}}\cdot \sqrt{2Nt}
  \leq \frac{1/t}{\sqrt{\log(N(t+1))}}\cdot\sqrt{2Nt}
  =  \frac{\sqrt{2N}}{\sqrt{t\log(N(t+1))}}
  \text.
  \eqlabel{eq:ps-varying-smoothing-rate-termii-0}
\)
Based on this we get
\(
  \sum_{1\leq t\leq T}\frac{\log\frac1{1-\eps_t}}{\log\frac1\alpha_t}
  \stackrel{\eqref{eq:ps-varying-smoothing-rate-termii-0}}\leq \sqrt{2N}\cdot \sum_{1\leq t\leq T}\frac1{\sqrt{t\log(N(t+1))}}
  &\stackrel{\text{\ref{it:ps-varying-smoothing-rate-termii-step0}}}\leq \sqrt{2N}\cdot\left[\frac1{\sqrt{\log(2N)}} + \sum_{1<t\leq T} \frac1{\sqrt{t\log t}}\right]\\
  &\stackrel{\text{\ref{it:ps-varying-smoothing-rate-termii-step1}}}\leq \sqrt{2N}\cdot\left[\frac1{\sqrt{\log(2N)}} + 8\sqrt{\frac T{\log T}}\right]
  \text,
  \eqlabel{eq:ps-varying-smoothing-rate-termii}
\)
where
\renewcommand\theenum{(\alph{enum})}
\begin{enum}
\item\label{it:ps-varying-smoothing-rate-termii-step0}
  we split off the sum's first term and used $\log(N(t+1))\geq\log t$, for $t>1$, and
\item\label{it:ps-varying-smoothing-rate-termii-step1}
  we applied \lemref{lem:erfi-sum} (note that $T\geq 2$).
\end{enum}

\block{Terms (iii), (iv) and (v).}
The remaining sums can be bounded from above by simple arithmetics,
\(
  \sum_{1\leq t\leq T}(D(u\parallel p_t)-D(u\parallel p_{t+1}))
  &= D(u\parallel p_1)-D(u\parallel p_{T+1})
  \leq D(u\parallel p)
  \text,
  \eqlabel{eq:ps-varying-smoothing-rate-termiii}
  \\
  \sum_{1\leq t\leq T} \log\frac1{1-\eps_t}
  &\stackrel{\eqref{eq:ps-varying-parameters-pmin}}\leq \sum_{1\leq t\leq T} \log\frac{t+1}t
  = \log(T+1)
  \text,
  \eqlabel{eq:ps-varying-smoothing-rate-termiv}
  \\
  \sum_{1\leq t\leq T} \log\frac1{\alpha_t}
  &=\sum_{1\leq t\leq T} \sqrt{\frac{\log(N/\eps_t)}{2Nt}}
  \stackrel{\eqref{eq:ps-varying-parameters-monotonicity}}\leq \sqrt{\frac{\log(N/\eps_T)}{2N}}\cdot\sum_{1\leq t\leq T} \frac1{\sqrt t}\\
  &\mathrel{\phantom=}\phantom{\sum_{1\leq t\leq T} \sqrt{\frac{\log(N/\eps_t)}{2Nt}}}
  \mathrel\leq \sqrt{\frac{2 (T+1) \log(N(T+1))}{N}}
  \text,
  \eqlabel{eq:ps-varying-smoothing-rate-termv}
\)
where for the last inequality in \eqref{eq:ps-varying-smoothing-rate-termv} we used $\sum_{1\leq t\leq T} \frac1{\sqrt t} \leq 1+\int_1^T \frac{\mathrm dt}{\sqrt t} \leq 2\sqrt{T}$.

\block{Competing with \pws.}
By plugging \eqref{eq:ps-varying-smoothing-rate-termi} and \eqref{eq:ps-varying-smoothing-rate-termii}\,-\,\eqref{eq:ps-varying-smoothing-rate-termv} into \eqref{eq:ps-varying-smoothing-rate-2} we get
\(
  &\ell_{\ps}(x_{1:T})-\ell_{\pws'}(x_{1:T})\\
  &\qquad\leq \sqrt{2N(T+1)\log(N(T+1))}\cdot\C_{\pws'} + \sqrt{2N}\cdot\left[\frac1{\sqrt{\log(2N)}} + 8\sqrt{\frac T{\log T}}\right]\\
  &\omit{\hfil$\displaystyle+\mkern6mu N D(u\parallel p) + N \log(T+1) + \sqrt{2N(T+1)\log(N(T+1))}$}
  \text.
  \eqlabel{eq:ps-varying-smoothing-rate-5}
\)
To end the proof we bound the \lhs\ of \eqref{eq:ps-varying-smoothing-rate-5} from below by \eqref{eq:ps-varying-smoothing-rate-1b} and the \rhs\ from above by \eqref{eq:ps-varying-smoothing-rate-1a} and rearrange.
\hfill\qed
\end{prf}

\fi

\end{document}